\documentclass[aps,prd,eqsecnum,preprintnumbers,showpacs]{revtex4}
\def\sqr#1#2{{\vcenter{\vbox{\hrule height.#2pt\hbox{\vrule
width.#2pt height#1pt \kern#1pt\vrule width.#2pt}\hrule height.#2pt}}}}
 
\usepackage{latexsym}
\usepackage{bm}
\usepackage{graphicx}
\begin{document}

\vspace*{-10.0mm} % for revtex
\thispagestyle{empty}

\rightline{\large\baselineskip20pt\rm\vbox to20pt{
\baselineskip14pt
\hbox{YITP-04-22}}}
%\vspace{1mm}

\vskip8mm
\begin{center}
{\large\bf
Linearized gravity on the de Sitter brane in the Einstein Gauss-Bonnet
theory}
\end{center}

\begin{center}
\large
Masato Minamitsuji$^{1,2}$
%\footnote{E-mail:masato@yukawa.kyoto-u.ac.jp}
and 
Misao Sasaki$^2$
%\footnote{E-mail:misao@yukawa.kyoto-u.ac.jp}}
\end{center}

\begin{center}
{\em
$^1$Department of Earth and Space Science, Graduate School of Science,
\\
Osaka University, Toyonaka 560-0043, Japan}
\end{center}

\begin{center}
{\em
$^2$Yukawa Institute for Theoretical Physics,
\\
 Kyoto University, Kyoto 606-8502, Japan}
\end{center}

\begin{abstract}
We investigate the linearized gravity on a single de Sitter brane
in the anti-de Sitter (AdS) bulk in the Einstein Gauss-Bonnet 
(EGB) theory. 
We find that the Einstein gravity is recovered for a high energy
brane, i.e., in the limit of the large expansion rate, $H\ell\gg 1$,
where $H$ is the de Sitter expansion rate
and $\ell$ is the curvature radius of the AdS bulk.
 We also show that, in the short distance limit 
$r\ll \min\{\ell,H^{-1}\}$,
the Brans-Dicke gravity is obtained, whereas in the large
distance limit $r\gg\max\{\ell,H^{-1}\}$,
a Brans-Dicke type theory is obtained for $H\ell=O(1)$, and
the Einstein gravity is recovered both for
 $H\ell\gg1$ and $H\ell\ll1$.
In the limit $H\ell\to0$, these results
smoothly match to the results known for the Minkowski brane.
\end{abstract}


\date{\today}

\maketitle

\makeatletter
\makeatother

\section{Introduction}

Recent progress in string theory suggests that our universe is not
a four-dimensional spacetime in reality, but is a four-dimensional
submanifold ``brane'' embedded in a higher dimensional spacetime
called ``bulk''. As a simple realization of this braneworld,
the model proposed by Randall and Sundrum (RS)~\cite{Randall:1999vf}  
has attracted much attention because of its interesting feature that
gravity is localized on the brane not because of compactification
but by warping of the extra dimension.
This model is a solution of the 5-dimensional Einstein equations
with a negative cosmological constant, where a Minkowski brane is 
embedded in the 5-dimensional anti-de Sitter (AdS) bulk.
The linear perturbation theory in the RS model reveals
that the Einstein gravity is realized on the brane in the large
distance limit. However, in the short distance limit, 
the gravity on the brane becomes essentially 5-dimensional,
which may be interpreted as due to large contribution of
the Kaluza-Klein corrections~\cite{Garriga:1999yh}.
Cosmological extension of this model, inclusion of black holes
and so on, have been discussed by various authors \cite{Maartens:2003tw}.

{}From the stringy point of view, it is plausible that there may exist
many fields and higher-order curvature corrections in addition to
the bulk cosmological constant.
In this paper, we consider the gravitational action
with the Gauss-Bonnet term added to the usual Einstein-Hilbert term.
This type correction appears as low energy corrections
in the perturbative approach to string theory,
and is a natural extension of the Einstein-Hilbert action from
4-dimensions to higher dimensions~\cite{Myers:yn, Nojiri:2001ae, Deruelle:2003ck}.
Cosmological braneworld models in the Einstein Gauss-Bonnet (EGB)
theory have been 
discussed in~\cite{Kim:2000je, Abdesselam:2001ff, Charmousis:2002rc, Gravanis:2002wy,
 Lidsey:2003sj, Dehghani:2004cf, Maeda:2003vq},
and black holes in the EGB theory have been discussed 
in~\cite{Crisostomo:2000bb, Cai:2001dz, Cvetic:2001bk, Cvitan:2002cs, Cvitan:2002rh, Clunan:2004tb}.

Recently, Deruelle and Sasaki showed that in the 
EGB theory, the linearized
gravitational force on the
Minkowski brane behaves like a 4-dimensional one even in the
short distance limit~\cite{Deruelle:2003tz}.
Then Davis showed that the Brans-Dicke gravity
\cite{Brans:sx} is realized on the Minkowski
 brane in the short distance limit \cite{Davis:2004xa}.
Although, the effective gravitational theory in
 the nonlinear regime is unknown at all, these results imply
that the experimental
constraint on the maximum size (curvature radius) of the extra-dimension 
is drastically relaxed when compared with the RS model in which
the size of the extra-dimension must be less than $\sim0.1\,$mm.
Thus, the EGB theory deserves more detailed
investigations from various aspects.

In this paper, as a step toward understanding cosmological
implications of the EGB theory, we investigate
the linear perturbation of a single de Sitter brane embedded in
the AdS bulk. 
This paper is organized as follows.
In Sec.~II, we describe our set-up in the EGB theory. 
We consider an AdS bulk with a single de Sitter brane as the background
spacetime.
In Sec.~III, we analyze the linear perturbation theory in the bulk and
on the de Sitter brane. 
In Sec.~IV, we discuss the effective gravity theory on the brane for
various limits.
In the limit $H\ell\gg 1$, where $H$ is the expansion rate of the
de Sitter brane and $\ell$ is the AdS curvature radius,
we find that the Einstein gravity with a cosmological constant is
recovered on the de Sitter brane.
We also show that the Brans-Dicke gravity
is obtained in the short distance limit, whereas 
in the large distance limit a Brans-Dicke type theory
is obtained for $H\ell=O(1)$ and the Einstein gravity both for 
$H\ell\gg 1$ and  $H\ell\ll 1$. Furthermore, it is shown that
the results for the Minkowski brane are recovered in the limit $H\ell \to 0$,
namely, the Brans-Dicke gravity in the short distance limit and
the Einstein gravity in the large distance limit~\cite{Davis:2004xa}.
In Sec.~V, we briefly summarize our results.
In Appendix~A, we review the results for the Minkowski
 brane~\cite{Davis:2004xa}.
In Appendix~B, we define harmonic functions on the de Sitter
spacetime that correspond to the Fourier modes in the
Minkowski spacetime.
In Appendix~C, we consider the case of two de Sitter branes
and show that there exists a tachyonic bound state mode
that makes the system unstable, just as in the Minkowski case
discussed in~\cite{Charmousis:2003sq}.

\section{Einstein Gauss-Bonnet braneworld}

We consider a braneworld in the EGB theory with a
cosmological constant. As usual, we assume the mirror symmetry
with respect to the brane. Then we may focus on one of the
two identical copies of the spacetime $M$ with the brane
as the boundary $\partial M$.
The action is given by~\cite{Myers:yn, Deruelle:2003ck};
\begin{eqnarray}
S&=&\int_{M} d^5x\sqrt{-g}\frac{1}{2\kappa_{5}^2}
 \Bigl[{}^{(5)}R-2\Lambda_{5}
+\alpha\Bigl({}^{(5)}R^2-4{}^{(5)}R_{ab}{}^{(5)}R^{ab}+
             {}^{(5)}R_{abcd}{}^{(5)}R^{abcd} \Bigr) 
  \Bigr]   \label{action2}
\\ \nonumber
&+& \int_{\partial M}d^4 x\sqrt{-q}
\Bigl[-\sigma+{\cal L}_{m}+\frac{1}{\kappa_{5}^2}
  \Bigl(K+2\alpha \Bigl(J-2G^{\mu}{}_{\nu}K^{\nu}{}_{\mu}
 \Bigr) \Bigr)        
 \Bigr],
\end{eqnarray}
where $\alpha$ is the coupling constant for the Gauss-Bonnet term 
which has dimensions $(\mathrm{length})^2$, $\Lambda_{5}$ is the negative
cosmological constant, $g_{ab}$ and $q_{\mu\nu}$ are the bulk and brane
metrics, respectively.
${\cal L}_{m}$ is the Lagrangian density of the matter on the brane,
and $\sigma$ is the brane tension.
The second term in the second line in Eq. (\ref{action2}) denotes the 
generalized Gibbons-Hawking term~\cite{Gibbons:mu} which is added to the
boundary action in order to obtain the well-defined boundary value
problem.
$K_{\mu\nu}$ is extrinsic curvature of the brane and 
\begin{eqnarray}
J^{\mu}{}_{\nu}
= -\frac{2}{ 3} K^{\mu}{}_{\rho}K^{\rho}{}_{\sigma} K^{\sigma}{}_{\nu}
+\frac{2}{3}KK^{\mu}{}_{\rho}K^{\rho}{}_{\nu}
+\frac{1}{3} K^{\mu}{}_{\nu}\Bigl(
 K^{\rho\sigma}K_{\rho\sigma}-K^2 \Bigr). 
\end{eqnarray} 
The Latin indices $\{a,b,\cdots\}$ and the Greek indices
$\{\mu,\nu,\cdots\}$ are used for tensors defined in the
bulk and on the brane, respectively.

Extremizing the action $S$ with respect to the bulk metric,
the vacuum bulk Einstein Gauss-Bonnet equation is obtained as
\begin{eqnarray}
&&{}^{(5)}G_{ab}+\Lambda_5 \,g_{ab}
+\alpha\Bigl[2\Bigl({}^{(5)}R_{a}{}^{cde}{}^{(5)}R_{bcde}
-2{}^{(5)}R^{cd}R_{acbd}-2{}^{(5)}R_{ac}{}^{(5)}R^{c}{}_{b}
 +{}^{(5)}R{}^{(5)} R_{ab}\Bigr)\label{fullEGB} 
\\ \nonumber
&&\hspace{2cm}
-\frac{1}{2}g_{ab}\Bigl({}^{(5)}R^2-4{}^{(5)}R_{cd}{}^{(5)}R^{cd}+
{}^{(5)}R_{cdef}{}^{(5)}R^{cdef} \Bigr)   \Bigr]=0\,.
\end{eqnarray}
The brane trajectory is determined by the junction condition
which is obtained by varying the action $S$ with respect to
the brane metric~\cite{Israel:rt, Davis:2002gn}:
\begin{eqnarray}
B^{\mu}{}_{\nu}
=K^{\mu}{}_{\nu}-K\delta^{\mu}{}_{\nu}
+4\alpha
\Bigl(\frac{3}{2}J^{\mu}{}_{\nu}-\frac{1}{2}J\delta^{\mu}{}_{\nu}
-P^{\mu}{}_{\rho\nu\sigma}K^{\rho\sigma} \Bigr)
=\frac{1}{2}\kappa_{5}^2T^{\mu}{}_{\nu},
\label{fulljunction}
\end{eqnarray}
where
\begin{eqnarray}
&&P_{\mu\rho\nu\sigma}
:=R_{\mu\rho\nu\sigma}
+\Bigl(R_{\mu\sigma}q_{\rho\nu}-R_{\rho\sigma}q_{\mu\nu}
+R_{\rho\nu}q_{\mu\sigma}-R_{\mu\nu}q_{\rho\sigma}\Bigr)
\\ \nonumber
&&\hspace{2cm}-\frac{1}{2}R
\Bigl(q_{\mu\sigma}q_{\rho\nu} -q_{\mu\nu}q_{\rho\sigma}\Bigr),
\end{eqnarray}
and $T_{\mu\nu}$ is the energy momentum tensor of the matter on
the brane, defined as
\begin{eqnarray}
\delta\Bigl(\sqrt{-q}{\cal L}_{m}\Bigr)
=-\frac{1}{2}\sqrt{-q}T_{\mu\nu}\delta q^{\mu\nu}.
\end{eqnarray}
Note that the extrinsic curvature here is the one for the vector
normal to $\partial M$ pointing outward from the side of $M$.

\section{de Sitter brane in the Einstein Gauss-Bonnet theory}

Let us consider a de Sitter brane in the AdS bulk in the EGB theory,
and investigate the linearized gravity on the de Sitter brane.

\subsection{de Sitter brane in the Einstein Gauss-Bonnet theory}

We take the Gaussian normal coordinate with respect to the brane,
and assume the bulk metric in the form~\cite{Garriga:1999bq},
\begin{eqnarray}
ds^2= dy^2+b^2(y)\gamma_{\mu\nu} dx^{\mu}dx^{\nu}\,,
\end{eqnarray}
where $\gamma_{\mu\nu}$ is the metric of the
4-dimensional de Sitter spacetime with $R(\gamma)=12H^2$.

The background Einstein Gauss-Bonnet equation is
\begin{eqnarray}
-3H^2+3b''b+3b'^2-12\alpha \frac{b''}{b} \bigl(b'^2-H^2\bigr) 
=-\Lambda_{5}b^2.
\end{eqnarray}
This has a solution,
\begin{eqnarray}
b(y)=H\ell\sinh(y/\ell)\,,
\end{eqnarray}
where $\ell$ is given by
\begin{eqnarray}
\frac{1}{\ell^2}=\frac{1}{4\alpha}
\Bigl(1\pm \sqrt{1+\frac{4\alpha \Lambda_{5}}{3} } \Bigr),
\label{AdS}
\end{eqnarray}
This agrees with the Minkowski
 brane case~\cite{Deruelle:2003tz, Cho:2001su}.  
Without loss of generality, we choose the location of
the de Sitter brane at
\begin{eqnarray}
b(y_{0})=1\,.
\end{eqnarray}
Thus $H$ is the expansion rate of the de Sitter brane.

\subsection{Bulk gravitational perturbations}

First we consider gravitational perturbations in the bulk.
We take the RS 
gauge~\cite{Randall:1999vf, Garriga:1999yh, Gen:2002rb},
\begin{eqnarray}
h_{55}=h_{5\mu}=0\,,
\quad
h^{\mu}{}_{\mu}=D_{\nu}h^{\nu}{}_{\mu}=0\,,
\end{eqnarray}
where $D_{\alpha}$ denotes the covariant derivative
with respect to $\gamma_{\mu\nu}$,
and the perturbed metric is given by 
\begin{eqnarray}
ds^2= dy^2+b^2(y)
\Bigl(\gamma_{\mu\nu} +h_{\mu\nu}\Bigr)dx^{\mu}dx^{\nu}\,.
\end{eqnarray}

The ($\mu$,$\nu$)-components of the linearized Einstein Gauss-Bonnet
equation are given by
\begin{eqnarray}
\Bigl(1-\bar{\alpha}\Bigr)
\Bigl[\frac{1}{\sinh^{4}(y/\ell)}\partial_{y}
\Bigl( \sinh^{4}(y/\ell) \partial_{y} \Bigr)
+\frac{1}{ (H\ell)^2 \sinh^2(y/\ell)}
\left(\Box_{4}-2H^2\right)
\Bigr]h_{\mu\nu}=0\,,
\label{EGB}
\end{eqnarray}
where 
\begin{eqnarray}
\bar\alpha=\frac{4\alpha}{\ell^2}\,,
\end{eqnarray}
and $\Box_4=D^\mu D_\mu$ is the d'Alembertian with respect to
$\gamma_{\mu\nu}$. Throughout this paper,
 we assume $\bar{\alpha} \neq 1$.

Equation~(\ref{EGB}) is separable. Setting
 $h_{\mu\nu}=\psi_p(y)Y^{(p,2)}_{\mu\nu}(x^\alpha)$, we obtain
\begin{eqnarray}
&& \Bigl[\frac{1}{\sinh^{4}(y/\ell)}  \partial_{y}
\Bigl( \sinh^{4}(y/\ell) \partial_{y} \Bigr)
+\frac{m^2}{\ell^2 \sinh^2(y/\ell)}
 \Bigr]\psi_{p} (y)=0, \label{separable}
 \\  
&&
\Bigl[\Box_4-(m^2+2)H^2\Bigr]Y^{(p,2)}_{\mu\nu}=0, 
\label{tensor}
\end{eqnarray}
where $p^2=m^2-9/4$ and $Y^{(p,2)}_{\mu\nu}$ are the tensor-type
tensor harmonics on the de Sitter spacetime
which satisfy the gauge condition~\cite{Gen:2000nu},
\begin{eqnarray}
Y^{(p,2)\mu}{}_{\mu}=D_{\nu} Y^{(p,2)\nu}{}_{\mu}=0.
\end{eqnarray}
The properties of these harmonics are discussed in Appendix B.

The equation~(\ref{separable}) is the same as that
for a massless scalar field in the bulk~\cite{Kobayashi:2000yh}. 
There exists a mass gap for the eigenvalue 
$0<m<3/2$~\cite{Garriga:1999bq}. 
There is a unique bound state at $m=0$, which gives $\psi_p(y)=$constant,
and it is called the zero mode.
For $m>3/2$, the mass spectrum is continuous and 
they are called the Kaluza-Klein modes.
The general solution is
\begin{eqnarray}
\psi_p(y)=\frac{1}{\sinh^{3/2}(y/\ell)}
\Bigl[A_p P^{ip}_{3/2}(\cosh(y/\ell))
  +  B_p Q^{ip}_{3/2}(\cosh(y/\ell))
\Bigr],
\end{eqnarray} 
where $P^{\mu}_{\nu}(z)$ and $Q^{\mu}_{\nu}(z)$ are the
associated Legendre functions of the 1st and 2nd kinds, respectively.

For $p^2>0$ ($m>3/2$), we choose those harmonic functions 
$Y^{(p,2)}_{\mu\nu}$ that behave as $e^{-ipt}$ in the limit
 $t\to\infty$. 
Then, assuming that there is no incoming wave from the past infinity $y=0$,
we find we should take $B_{p}=0$. In fact, the asymptotic behavior of
$P^{ip}_{3/2}$ for $y\to 0$ is \cite{Bateman:1969ku}
\begin{eqnarray}
\frac{1}{\sinh^{3/2}(y/\ell)}
P^{ip}_{3/2} (\cosh(y/\ell))
&&\mathop{\longrightarrow}\limits_{y\to0}  
\frac{2^{ip}}{\Gamma(1-i p)}
\Bigl(\frac{\sinh(y/\ell)}{\cosh(y/\ell) }\Bigr)^{-ip-3/2}
\simeq
\frac{2^{ip}} {\Gamma(1-i p)}
 \Bigl(\frac{y}{\ell}\Bigr)^{-3/2}
e^{-ip \ln(y/\ell) }\,,
\end{eqnarray}
which guarantees the no incoming wave (i.e., retarded)
boundary condition.
Thus the bulk metric perturbations are constructed by
\begin{eqnarray}
h_{\mu\nu}=
\oint_C dp \, \psi_p(y)Y^{(p,2)}_{\mu\nu}(x^{\mu}),
\end{eqnarray}
where the contour of integration $C$ is chosen on the complex 
$p$-plane such that it runs from
$p=-\infty$ to $p=\infty$ and covers the bound state pole
at $p=3i/2$ below the contour~\cite{Himemoto:2001hu}.

\subsection{Linearized effective gravity on the brane}

We now investigate the effective gravity on the brane. The position of the brane in the coordinate system is displaced in
general as
\begin{eqnarray}
y=y_{0}-\ell\varphi(x^{\mu}), 
\end{eqnarray}
where the second term in the right-hand-side describes
the brane bending~\cite{Deruelle:2003tz, Gen:2000nu}. 
The induced metric on the brane is given by
\begin{eqnarray}
ds^2\Big|_{(4)}
=\Bigl(\gamma_{\mu\nu}+\tilde h_{\mu\nu}\Bigr)dx^\mu dx^\nu\,;
\quad
\tilde h_{\mu\nu}
=h_{\mu\nu}-2\coth(y_0/\ell)\,\varphi \gamma_{\mu\nu}\,.
\label{induced}
\end{eqnarray}
The extrinsic curvature on the brane is given by
\begin{eqnarray}
K^{\mu}{}_{\nu}=\frac{1}{\ell}
  \coth(y_{0}/\ell)\delta^{\mu}{}_{\nu}
+\frac{1}{2} h^{\mu}{}_{\nu,y}
+\ell \Bigl(D^{\mu}D_{\nu}+ H^2  \delta^{\mu}{}_{\nu}\Bigr)\varphi
\,.
\end{eqnarray}

We consider the junction condition~(\ref{fulljunction}).
The background part gives the relation between the
brane tension and the location of the brane,
\begin{eqnarray}
\kappa_{5}^2\sigma
 =\frac{6}{\ell}\coth(y_0/\ell)
\left(1-\frac{\bar{\alpha}}{3}
+\frac{2\bar\alpha}{3\sinh^2(y_0/\ell)}
\right)\,,
\label{bgsigma}
\end{eqnarray}
where
\begin{eqnarray}
\coth(y_0/\ell)=\sqrt{1+(H\ell)^2}\,,
\quad
\sinh(y_0/\ell)=\frac{1}{H\ell}\,.
\end{eqnarray}
In the limit $H\ell\ll 1$, Eq.~(\ref{bgsigma}) reduces to 
the Minkowski tension,
\begin{eqnarray}
\kappa_{5}^2\sigma \simeq \frac{6}{\ell}
 \Bigl(1-\frac{1}{3}\bar{\alpha} \Bigr).
\end{eqnarray}
The perturbative part of the junction condition
gives 
\begin{eqnarray}
\Bigl(1+\beta\Bigr)
\Bigl(D_{\mu}D_{\nu}-\Box_{4}\gamma_{\mu\nu} 
-3H^2\gamma_{\mu\nu}\Bigr)\varphi
+\frac{1}{2\ell}\Bigl(1-\bar{\alpha}\Bigr) h_{\mu\nu,y}
-\frac{1}{2}\bar{\alpha}\coth(y_0/\ell)
\Bigl(\Box_{4}-2H^2 \Bigr)  h_{\mu\nu}
=\frac{\kappa_{5}^2}{2\ell}\,S_{\mu\nu}\,,
 \label{pertjunc}
\end{eqnarray}
where
\begin{eqnarray}
\beta :=
\frac{\cosh^2(y_0/\ell)+1}{\sinh^2(y_0/\ell)}\,\bar{\alpha}
=\left(2\coth^2(y_0/\ell)-1\right)\bar\alpha
=\left(2(H\ell)^2+1\right)\bar\alpha\,. 
\label{betadef}
\end{eqnarray}

The trace of Eq. (\ref{pertjunc}) gives the equation to determine the 
brane bending as
\begin{eqnarray}
\Bigl(\Box_{4} +4H^2\Bigr) \varphi=
-\frac{\kappa_5^2}{6(1+\beta)\ell}\,S\,,
\label{bendion}
\end{eqnarray}
where $S=S^\mu{}_\mu$.
Note that the field $\varphi$ seems to be tachyonic, with mass-squared
given by $-4H^2$. However, in the case of a de Sitter brane
in the Einstein gravity, there was a similar equation for the
brane bending, but it was found to be non-dynamical~\cite{Gen:2000nu}.
We shall see below that the situation is
quite similar in the present case of the EGB theory.

To find the effective gravitational equation on the brane,
we manipulate as follows. 
Using the expression for the induced metric on the brane,
Eq.~(\ref{induced}), 
the perturbation of the brane Einstein tensor is given by
\begin{eqnarray}
\delta G_{\mu\nu}[\tilde h]
&=&-\frac{1}{2}\Box_{4} h_{\mu\nu}-2H^2h_{\mu\nu}
+2\coth(y_0/\ell)\Bigl[D_{\mu}D_{\nu}
-\Box_{4}\,\gamma_{\mu\nu}\Bigr] \varphi
\nonumber
\\
&=&-3H^2\Bigl(h_{\mu\nu}-2\coth(y_0/\ell)\gamma_{\mu\nu}\varphi\Bigr)
\nonumber\\
&&\quad
-\frac{1}{2}\Bigl(\Box_{4} -2H^2\Bigr)h_{\mu\nu}
+2\coth(y_0/\ell)
\Bigl[D_{\mu}D_{\nu}-\Box_{4}\,\gamma_{\mu\nu}
-3H^2\gamma_{\mu\nu}\Bigr] \varphi
\,.
\label{Einstein}
\end{eqnarray}
Using the perturbed junction condition~(\ref{pertjunc})
we can eliminate the term involving $\varphi$ from the above
equation to obtain
\begin{eqnarray}
\delta G_{\mu\nu}[\tilde h]
+3H^2\tilde h_{\mu\nu}
=-\frac{1-\bar{\alpha}}{ 2\bigl( 1+\beta \bigr)}
\Bigl(\Box_4-2H^2\Bigr)h_{\mu\nu}
-\frac{1-\bar{\alpha}}{\ell\bigl(1+\beta \bigr)}
\coth(y_0/\ell)h_{\mu\nu,y}
+\frac{\kappa_{5}^2}{\ell}
\frac{\coth(y_0/\ell)}{\bigl(1+\beta \bigr)}\,S_{\mu\nu}.  
\label{Ejunk}
\end{eqnarray}
Eliminating the term proportional to $(\Box_4-2H^2)h_{\mu\nu}$ from
Eqs.~(\ref{Einstein}) and (\ref{Ejunk})
we obtain
\begin{eqnarray}
\delta G_{\mu\nu}[\tilde h]
+ 3H^2\tilde h_{\mu\nu}
&=&\frac{\kappa_{5}^2\tanh(y_0/\ell)}{2\ell\,\bar\alpha}\,S_{\mu\nu}
-\frac{1-\bar{\alpha}}{\bar{\alpha}}\tanh(y_0/\ell)
 \Bigl( D_{\mu}D_{\nu} -\gamma_{\mu\nu} \Box_{4}
 -3H^2 \gamma_{\mu\nu} \Bigr)\varphi
\nonumber\\
&&\qquad
-\frac{1-\bar{\alpha}}{2\ell\bar\alpha}\tanh(y_0/\ell)
h_{\mu\nu,y}\,.
 \label{branefluc}
\end{eqnarray}
Together with Eq.~(\ref{bendion}),
this may be regarded as an effective gravitational equation
on the brane. The effect of the bulk gravitational field
is contained in the last term proportional to $h_{\mu\nu,y}$.

\subsection{Harmonic decomposition}

Using the harmonic functions defined in Appendix B,
we may obtain a closed (integro-differential) system
on the brane. We decompose the perturbations on the brane 
as
\begin{eqnarray}
&&S_{\mu\nu}= S^{(0)}_{\mu\nu}+S^{(2)}_{\mu\nu}\,;
\quad S^{(0)}_{\mu\nu}=\int^{\infty}_{-\infty}dp\,      
           \Bigl(S_{(p,0)}Y^{(p,0)}_{\mu\nu}\Bigr)\,,
\quad S^{(2)}_{\mu\nu}=\int^{\infty}_{-\infty}dp\,      
           \Bigl(S_{(p,2)}Y^{(p,2)}_{\mu\nu}\Bigr),
\nonumber \\
&&\varphi=\int^{\infty}_{-\infty}dp\,
        \varphi_{(p)} Y^{(p,0)},
\nonumber\\
&& h_{\mu\nu}=\int^{\infty}_{-\infty}dp\, h_{(p)}Y^{(p,2)}_{\mu\nu}\,,
\end{eqnarray}
where $Y^{(p,0)}$ are the scalar harmonics and 
$Y^{(p,0)}_{\mu\nu}$ are the scalar-type tensor harmonics
given in terms of $Y^{(p,0)}$, as defined in Appendix B.
Note that, because of the energy-momentum conservation, $D^\mu S_{\mu\nu}=0$,
there is no contribution from the vector-type tensor harmonics
which do not satisfy the divergence free condition.
If a bound state exists, we have to deform the contour of integration
so that the corresponding pole is covered, as mentioned at the end
of subsection~B.

With the above decomposition, the metric perturbation on the brane
$\tilde h_{\mu\nu}$ given by Eq.~(\ref{induced})
 consists of the isotropic scalar-type
part and tensor-type part.
The scalar-type part is determined by Eq. (\ref{bendion}),
which gives
\begin{eqnarray}
\varphi_{(p)}
=\frac{\kappa_{5}^2}{2(1+\beta)\ell}\, N_{p}\,S_{(p,0)}
=\frac{\kappa_{5}^2}{ 2(1+\beta)\ell}\,
  \frac{1}{ \sqrt{3}\sqrt{\bigl(p^2+\frac{21}{ 4}\bigr)
          \bigl(p^2+\frac{25}{4}\bigr)}H^2}\,S_{(p,0)}\,,
\label{fourierbendion}
\end{eqnarray}
where $N_p$ is the normalization factor for the harmonics defined in
Appendix B. We see that the propagator part of the above
(i.e., the coefficient of $S_{(p,0)}$) do not contain
the pole at $p=(5/2)i$ which would corresponds to
the tachyonic mode with mass-squared $-4H^2$.
Instead, it becomes a branch point and a branch cut
appears between the points $ p=(\sqrt{21}/{2})i$ and
$p=(5/2)i$. Thus we find the tachyonic mode is absent
and there is no instability associated with the brane bending
due to the matter source on the brane.

Before we proceed, it is useful to note the equation,
\begin{eqnarray}
\Bigl(D_{\mu}D_{\nu} -\gamma_{\mu\nu}\Box_{4}-3H^2 \gamma_{\mu\nu} 
\Bigr)\varphi=\frac{\kappa_5^2}{2(1+\beta)\ell}\,S^{(0)}_{\mu\nu}\,,
\label{scalarS}
\end{eqnarray}
which directly follows from Eq.~(\ref{fourierbendion}) and the
definition of the scalar-type tensor harmonics $Y^{(p,0)}_{\mu\nu}$.

There is a free
propagating tachyonic mode corresponding to the homogeneous solution of 
Eq.~(\ref{bendion}), which does not couple to either the scalar or
tensor-type matter perturbations on the brane.
However, we shall argue in the next subsection
that the mode that corresponds to
the exponential growth of the perturbation is unphysical,
namely, the only physical mode associated with this tachyonic
mode is exponentially decaying with time.

The traceless part of Eq.~(\ref{pertjunc}) gives
\begin{eqnarray}
h_{(p)}(y_0)
= -\frac{1}{(ip+3/2)}\,
\frac{\ell^2\,\sinh(y_0/\ell)\,P^{ip}_{3/2}(z_0)}
{\bigl(1-\bar{\alpha}\bigr) P^{ip}_{1/2}(z_0)
+ \bar{\alpha}\bigl(-ip+3/2\bigr)(H\ell)^2\cosh(y_0/\ell)
 P^{ip}_{3/2}(z_0)}\,\frac{\kappa_{5}^2}{\ell}\,S_{(p,2)}\,,
\label{fouriermetric}
\end{eqnarray}
where $z_0=\cosh(y_0/\ell)$.
This shows that the harmonic component of the tensor-type
metric perturbations on the brane has a simple pole at $p=(3/2)i$
on the complex $p$-plane, which corresponds to the zero mode.

For convenience, we also write down the $y$-derivative of
$h_{(p)}$,
\begin{eqnarray}
\frac{1}{\ell}\partial_y h_{(p)}(y_0)=
\frac{P^{ip}_{1/2}(z_0)}{(1-\bar\alpha)P^{ip}_{1/2}(z_0)+
\bar\alpha(-ip+3/2)(H\ell)^2\cosh(y_0/\ell)P^{ip}_{3/2}(z_0)}
\,\frac{\kappa_{5}^2}{\ell}\,S_{(p,2)}\,.
\label{fourierhd}
\end{eqnarray}
Then, Eqs~(\ref{bendion}), (\ref{branefluc}) and (\ref{fourierhd})
constitute the effective gravitational equations on the brane
that form a closed set of integro-differential equations.

\subsection{Source-free tachyonic mode}

Now, we discuss the source-free tachyonic mode on the 
brane~\cite{Gen:2002rb}.
This mode corresponds to the homogeneous solution of
Eq.~(\ref{bendion}), so does not couple to the matter 
perturbations on the brane.

On the complex $p$-plane, the solution corresponds to the pole 
at $p=(5/2)i$. Thus, the solution is given by
\begin{eqnarray}
\varphi=\varphi_{(5i/2)}Y^{(5i/2,0)}. 
\end{eqnarray}
For this mode, the junction condition~(\ref{pertjunc}) tells us
that it is associated with a non-vanishing $h_{\mu\nu}$.
The solution in the bulk is given by~\cite{Gen:2002rb} 
\begin{eqnarray}
h_{\mu\nu}=\phi(y){\cal L}_{\mu\nu} \varphi, \quad
{\cal L}_{\mu\nu}= D_{\mu}D_{\nu}+H^2\gamma_{\mu\nu}. 
\end{eqnarray}
This satisfies the transverse-traceless condition and
\begin{eqnarray}
\Bigl(\Box_{4} -4H^2\Bigr)h_{\mu\nu}=0.
\end{eqnarray}
Thus, this mode falls within the mass gap between $m=0$ and $3/2$,
with the mass $mH=\sqrt{2}H$.

Let us first analyze the behavior of the function $\phi(y)$.
It should satisfy Eq.~(\ref{EGB}), which becomes
\begin{eqnarray}
\Bigl[ \frac{1}{\sinh^4(y/\ell)} \partial_{y}
 \Bigl(\sinh^4(y/\ell)\partial_{y}\Bigr) 
+\frac{2}{\ell^2 \sinh^2(y/\ell)} \Bigr]\phi(y)=0.
\end{eqnarray}
The general solution is given by
\begin{eqnarray}
&&\phi(y)=c_{1} \phi_{1}(y)+c_{2} \phi_{2}(y)\,;
\nonumber
\\
&&\phi_{1}(y)= \coth(y/\ell),\quad
\phi_{2}(y)= 1+\coth^2(y/\ell),
\end{eqnarray}
where the coefficients $c_{1}$ and $c_{2}$ are related 
through the junction condition~(\ref{pertjunc}) as
\begin{eqnarray}
1-\frac{1}{2}H^2\,c_1-H^2\coth(y_0/\ell)
\frac{1+\bar{\alpha}\coth^2(y_0/\ell)}{1+\beta}\,c_2=0. 
\label{coef} 
\end{eqnarray}
As readily seen, this mode diverges badly as $y\to0$.
Therefore, the regularity condition at $y=0$ will eliminate
this mode. Nevertheless, since its effect on the brane
seems non-trivial, it is interesting to see 
the physical meaning of it.

We note that $\phi_1$ is a gauge mode.
This can be checked by calculating the projected Weyl tensor 
$E_{\mu}{}_{\nu}:={}^{(5)}C_{n\mu n\nu}$~\cite{Shiromizu:1999wj,Maeda:2003vq}
which is gauge-invariant. We find that only the coefficient $c_2$
survives:
\begin{eqnarray}
E^{\mu}{}_{\nu}(y,x^\alpha)
=-\frac{c_2}{\ell^2\sinh^4(y/\ell)}
{\cal L}^{\mu}{}_{\nu}\,\varphi(x^\alpha). 
\label{Emn}
\end{eqnarray}
This means that the junction condition~(\ref{coef}) does not fix
the physical amplitude $c_2$. It just fixes the gauge amplitude
$c_1$. 

To understand the physical meaning of this mode, it is
useful to analyze the temporal behavior the projected Weyl tensor.
For simplicity, let us consider a spatially homogeneous solution
for $\varphi$. Choosing the spatially closed chart for the de Sitter
brane, for which the scale factor is given by $a(t)=H^{-1}\cosh(Ht)$,
we find
\begin{eqnarray}
\varphi=
C_1 \frac{P^{5/2}_{1/2}(\tanh(Ht))}{\cosh^{3/2}(Ht)}
+C_2 \frac{P^{-5/2}_{1/2}(\tanh(Ht))}{\cosh^{3/2}(Ht)}
\mathop\sim\limits_{t\to\infty} \tilde C_1\, e^{Ht}+\tilde C_2 \,e^{-4Ht}\,,
\end{eqnarray}
where $\tilde C_1$ and $\tilde C_2$ differ from $C_1$ and $C_2$,
respectively, by unimportant numerical factors.
We see that the solution associated with $C_1$ is the one that
shows instability.
If we insert this solution to Eq.~(\ref{Emn}), however, this unstable
solution disappears. In fact, we obtain
\begin{eqnarray}
E^{t}{}_{t}\simeq 
\frac{15H^2\tilde{C}_2\, c_2}{\ell^2\sinh^{4}(y/\ell)e^{4Ht} }
\simeq
\frac{15(H\ell)^2\tilde{C}_2\, c_2}{16(H\ell)^4\sinh^{4}(y/\ell)a^{4}(t)}\,.
\label{ett1}
\end{eqnarray}
We note that $E^{t}{}_{t}$ on the brane decays as $1/a^4(t)$. This is exactly
what one expects for the behavior of the so-called dark radiation.
We also note that, although $E_{\mu\nu}$ does not vanish for spatially
inhomogeneous modes, they decay as $1/a^3(t)$~\cite{Gen:2002rb},
giving no instability to the brane. 

In the Einstein case, the dark radiation term appears if there exists a
black hole in the bulk. This is also true in the EGB case. 
There also exists a spherically symmetric black hole solution in the
EGB theory~\cite{Crisostomo:2000bb, Cai:2001dz, Cvetic:2001bk, Cvitan:2002cs, Cvitan:2002rh, Clunan:2004tb}.
The metric is given by
\begin{eqnarray}
ds^2=-f(R)dT^2
    +\frac{dR^2}{f(R)} +R^2d\Omega_{(3)}^2;\quad
f(R)=1+\frac{R^2}{4\alpha} 
\left(1-\sqrt{1+\frac{16\alpha \mu}{3 R^4}
 +\frac{4}{3}\alpha \Lambda_5 }\right),
\end{eqnarray} 
where $\mu=\kappa_{5}^2 M/(2\pi^2)$ and $M$ is the mass of the
black hole.
For this solution, the projected Weyl tensor is given by
\begin{eqnarray}
E^{t}{}_{t}
= \frac{\mu}{R^4}
\Bigl(1+\frac{4}{3}\alpha \Lambda_{5}
  +\frac{16\alpha\mu}{3R^4}\Bigr)^{-3/2}
\Bigl(1+\frac{4}{3}\alpha \Lambda_{5}
  +\frac{16\alpha\mu}{9R^4}\Bigr)
\simeq \frac{\mu}{R^4}
\Bigl(1+\frac{4}{3}\alpha \Lambda_{5}\Bigr)^{-1/2},
\label{ett2}
\end{eqnarray}
for $R\gg (\alpha \mu)^{\frac{1}{4}}$ \cite{Maeda:2003vq}.
Comparing Eq. (\ref{ett1}) with Eq. (\ref{ett2}), with the
identification $R={\ell}\sinh(y/\ell)\cosh(Ht)$,
we find
\begin{eqnarray}
c_2\, \tilde C_2
 \simeq 
%-
 \frac{16\mu}{15(H\ell)^2}
\Bigl(1+\frac{4}{3}\alpha \Lambda_{5}\Bigr)^{-1/2}.
\end{eqnarray}
Thus the solution that decays exponentially in time
corresponds to adding a small black hole in the
 bulk~\cite{Yoshiguchi:2004cb}. 

In the two-brane system, the mode discussed here corresponds to the
radion, which describes the relative displacement of the
 branes~\cite{Gen:2000nu, Gen:2002rb}.   
As the case of the Einstein gravity, the radion mode is truly tachyonic.
However, for the EGB theory,
there is a tachyonic bound state mode other than the
radionic instability, as in the limit of the Minkowski
 brane~\cite{Charmousis:2003sq}, 
as discussed explicitly in Appendix C.
This renders the two-brane system physically unrealistic
in the EGB theory.

\section{Linearized gravity in limiting cases}

In this section, we discuss the effective gravity
on the brane in various limiting cases. We find the effective
gravity reduces to 4-dimensional theories in all the limiting
cases.

\subsection{High energy brane: $H\ell\gg1$}

For a high energy brane, i.e., $H\ell \gg 1$ limit,
we have $\tanh(y_0/\ell)\simeq 1/(H\ell)$ and $\beta\simeq 2(H\ell)^2$.
We assume that matter perturbations on the brane are dominated by the
modes $p \sim O(1)$. Namely, we consider the case $H\ell\gg p$.
Then, from Eq.~(\ref{scalarS}) and Eq.~(\ref{fourierhd}), we find
that the second and the third terms in the right-hand-side of
Eq.~(\ref{branefluc}) are suppressed by $1/(H\ell)^2$
relative to the first term,
\begin{eqnarray}
&&\delta G_{\mu\nu}[\tilde h]+3H^2\tilde h_{\mu\nu}
=\frac{\kappa_{5}^2\tanh(y_{0}/\ell)}{2\ell\bar\alpha}
\Bigl(S_{\mu\nu}+O\bigl((H\ell)^{-2}\bigr)\Bigr)
\,.
\end{eqnarray}
Thus, we obtain Einstein gravity with the
cosmological constant $3H^2$, with the gravitational constant
$G_4$ given by
\begin{eqnarray}
8\pi G_{4}=\frac{\kappa_5^2}{2\ell\,\bar{\alpha}}\tanh(y_{0}/\ell)
\approx\frac{\kappa_{5}^2}{ 2(H\ell) \bar{\alpha} \ell}.
\end{eqnarray}  
The terms we have neglected give the low energy non-local
corrections:
\begin{eqnarray}
\Bigl(\delta G_{\mu\nu}[\tilde h]\Bigr)_{\mathrm{corr},H\ell}
&=& -\frac{\kappa_{5}^2 \bigl(1-\bar{\alpha}\bigr)}{2\ell\bar{\alpha}}
   \tanh(y_0/\ell)
 \nonumber
\\
&&\times
\int^{\infty}_{-\infty}
dp\, 
 \Biggl\{ Y^{(p,2)}_{\mu\nu}  S_{(p,2)}               
\left[
\frac{P^{ip}_{1/2} (z_0)}
{\bigl(1-\bar{\alpha}\bigr)P^{ip}_{1/2} (z_0)  
 + \bar{\alpha} \bigl(-ip+3/2\bigr) (H\ell)^2\cosh(y_0/\ell)
P^{ip}_{3/2}(z_0) }
\right]
\nonumber
\\
&&\qquad\qquad
+Y^{(p,0)}_{\mu\nu}S_{(p,0)} 
   \frac{1}{ (1+\beta)} \Biggr\}.    
\end{eqnarray}

\subsection{Short and large distance limits}

In order to discuss short and large distance limits,
it is convenient to start from the expression~(\ref{Einstein}) 
for the perturbed Einstein tensor, and Eq.~(\ref{scalarS})
which relates the brane bending scalar $\varphi$ to the scalar part
of the energy momentum tensor $S^{(0)}_{\mu\nu}$. 
Let us recapitulate these expressions:
\begin{eqnarray}
&&\delta G_{\mu\nu}[\tilde h]
+3H^2\tilde h_{\mu\nu}
=2\coth(y_0/\ell)
\Bigl(D_{\mu}D_{\nu} -\gamma_{\mu\nu}\Box_{4}-3H^2 \gamma_{\mu\nu} 
 \Bigr)\varphi
-\frac{1}{2}\Bigl( \Box_{4}-2H^2 \Bigr) h_{\mu\nu}\,,
\label{Etensor}
\\
&&\Bigl(D_{\mu}D_{\nu} -\gamma_{\mu\nu}\Box_{4}-3H^2 \gamma_{\mu\nu} 
\Bigr)\varphi=\frac{\kappa_5^2}{2(1+\beta)\ell}\,S^{(0)}_{\mu\nu}\,.
\label{ddvarphi}
\end{eqnarray}

\par
\penalty-150
\vspace{5mm}
\noindent
{\bf 1. Short distance limit: $\bm{r\ll\min\{\ell,H^{-1}\}}$}
\vspace{3mm}

For the short distance limit $p\to \infty$, using Eq. (\ref{fouriermetric}),
we find
\begin{eqnarray}
&&-\frac{1}{2}\Bigl( \Box_{4}-2H^2 \Bigr) h_{\mu\nu}
\nonumber
\\ 
&&=
\frac{\kappa_{5}^2}{2\ell}\int^{\infty}_{-\infty}dp\,
 Y^{(p,2)}_{\mu\nu} S_{(p,2)}
\frac{(H\ell)\,(-i p+3/2)
  P^{ip}_{3/2}(z_0)/ 
   P^{ip}_{1/2}(z_0)}
{ \bigl(1-\bar{\alpha}\bigr)
  +\bar{\alpha}(H\ell)^2\cosh(y_0/\ell)\bigl(-ip+3/2 \bigr)
 P^{ip}_{3/2}(z_0)
 /  P^{ip}_{1/2}(z_0)}
\nonumber
\\
&&
\mathop{\longrightarrow}\limits_{p\to \infty}
\frac{\kappa_{5}^2}{2\ell\bar{\alpha}} \tanh(y_0/\ell)
\int^{\infty}_{-\infty}dp\, Y^{(p,2)}_{\mu\nu} S_{(p,2)}
%=\frac{\kappa_{5}^2}{\ell} \frac{\coth(y_0/\ell)}{\bar{\alpha}+\beta} 
%\int^{\infty}_{-\infty}dp\, Y^{(p,2)}_{\mu\nu} S_{(p,2)}
\,.
\label{shorthmn}
\end{eqnarray}
Also, using Eq.~(\ref{ddvarphi}), we manipulate as
\begin{eqnarray}
&&2\coth(y_0/\ell) 
\Bigl(D_{\mu}D_{\nu} -\gamma_{\mu\nu}\Box_{4}-3H^2 \gamma_{\mu\nu} 
\Bigr)\varphi 
\nonumber
\\
&& =\frac{\kappa_{5}^2}{2\ell\bar\alpha}\tanh(y_0/\ell)
\int^{\infty}_{-\infty}dp\,
S_{(p,0)} Y^{(p,0)}_{\mu\nu}
-\frac{1-\bar{\alpha}}{\bar{\alpha}} \tanh(y_0/\ell) 
\Bigl(D_{\mu}D_{\nu} -\gamma_{\mu\nu}\Box_{4}-3H^2 \gamma_{\mu\nu} 
\Bigr)\varphi\,, 
\label{shortphi}
\end{eqnarray}
where we have used an identity,
\begin{eqnarray}
2\coth(y_0/\ell)
&=&2\coth(y_0/\ell)-\frac{1+\beta}{\bar\alpha}\tanh(y_0/\ell)
+\frac{1+\beta}{\bar\alpha}\tanh(y_0/\ell)
\nonumber
\\
&=&-\frac{1-\bar\alpha}{\bar\alpha}\tanh(y_0/\ell)
+\frac{1+\beta}{\bar\alpha}\tanh(y_0/\ell)\,,
\end{eqnarray}
which follows from the definition of the parameter
 $\beta$, Eq.~(\ref{betadef}).

Substituting Eqs.~(\ref{shorthmn}) and (\ref{shortphi}) in
Eq.~(\ref{Etensor}), the linearized gravity on the brane 
at short distances becomes
\begin{eqnarray}
\delta G_{\mu\nu}[\tilde h]+3H^2\tilde h_{\mu\nu}
=\frac{\kappa_{5}^2}{2\ell\bar\alpha}\tanh(y_0/\ell)\,S_{\mu\nu}
-\frac{(1-\bar\alpha)}{\bar{\alpha}}\tanh(y_0/\ell)
\Bigl(D_{\mu}D_{\nu}-
\Box_{4}\gamma_{\mu\nu}-3H^2\gamma_{\mu\nu}\Bigr)\varphi\,,
\label{shortlimit}
\end{eqnarray}
with
\begin{eqnarray}
\Bigl(\Box_{4}+4H^2\Bigr)\varphi
= -\frac{\kappa_{5}^2}{6(1+\beta)\ell}\,S\,.
\end{eqnarray}
This is a scalar-tensor type theory. 

Interestingly, the scalar field $\varphi$ which
describes the brane bending degree of freedom
turns to be dynamical. As we have
seen in the previous subsection, there is no intrinsically dynamical
mode associated with the brane bending. Therefore, this emergence
of a dynamical degree of freedom is due to an accumulative
effect of the whole Kaluza-Klein modes, like a collective mode.
 Furthermore,
because of the tachyonic mass, the system appears to be unstable.
However, this is not the case. Since we have taken
the limit $p\to\infty$, all the perturbations have energy much larger
than $H$, and
the tachyonic mass-squared $-4H^2$ is completely negligible.
In other words, the spacetime appears to be flat at sufficiently
short distance scales.

We can rewrite Eq.~(\ref{shortlimit}) in the form,
\begin{eqnarray}
\delta G_{\mu\nu}[\tilde h]+\Lambda_{4} \tilde{h}_{\mu\nu}
&=&\frac{1}{\Phi_{0}}\Bigl(D_{\mu}D_{\nu}-
\Box_{4}\gamma_{\mu\nu}-3H^2\gamma_{\mu\nu}\Bigr)\delta \Phi
+\frac{8\pi G_{4}}{ \Phi_{0}}S_{\mu\nu}, \label{fluctuation}
\nonumber   
\\
\Bigl(\Box_{4}+4H^2\Bigr)\delta\,\Phi
&=& \frac{8\pi G_{4}}{ 3+2\omega }S \,, 
\end{eqnarray} 
with the identifications,
\begin{eqnarray}
\frac{8\pi G_{4}}{ \Phi_{0}}
=\frac{\kappa_5^2}{2\ell\bar{\alpha}}\tanh(y_0/\ell)\,,
\quad
\frac{\delta\Phi}{\Phi_{0}}=
     -\frac{1-\bar{\alpha}}{ \bar{\alpha}}\tanh(y_0/\ell)\varphi,
\quad
\omega=\frac{3\bar{\alpha}}{ 1-\bar{\alpha}}\coth^2(y_0/\ell)\,,
\quad
\Lambda_{4}=3H^2.
\end{eqnarray}
Neglecting the tachyonic mass of $\delta\Phi$, as justified above,
this is the linearized Brans-Dicke gravity with a cosmological
constant~\cite{Brans:sx}.
For $H\ell\ll1$, we have $\tanh(y_0/\ell)\simeq\coth(y_0/\ell)\simeq 1$.
Then
\begin{eqnarray}
\frac{8\pi G_{4}}{\Phi_{0}}\simeq 
\frac{\kappa_{5}^2}{2 \bar{\alpha}\ell},\quad
\frac{\delta\Phi}{\Phi_{0}}\simeq 
 -\frac{1-\bar{\alpha}}{\bar{\alpha}}
\varphi, \quad
\omega \simeq \frac{3\bar{\alpha}}{1-\bar{\alpha}}.
\end{eqnarray} 
This is in agreement with the Minkowski brane case
investigated recently~\cite{Davis:2004xa}.

The corrections are written as
\begin{eqnarray}
&&\Bigl(\delta G_{\mu\nu}[\tilde h] \Bigr)_{\mathrm{corr}, p\gg1}
\nonumber 
\\ 
&&\quad
= -\frac{\kappa_{5}^2}
         {2\ell\bar{\alpha}}\tanh(y_0/\ell)
\int^{\infty}_{-\infty} dp\,Y^{(p,2)}_{\mu\nu}  S_{(p,2)} 
\Biggl[ \frac{ \bigl(1-\bar{\alpha}\bigr)P^{ip}_{1/2} (z_0)}
{\bigl(1-\bar{\alpha}\bigr)P^{ip}_{1/2} (z_0)  
 + \bar{\alpha} (H\ell)^2 \bigl(-ip+3/2\bigr) \cosh(y_0/\ell)
P^{ip}_{3/2}(z_0) }
\Biggr] .
\end{eqnarray}

\par
\penalty -150
\vspace{5mm}
\noindent
{\bf 2. Large distance limit: $\bm{r\gg\max\{\ell,H^{-1}\}}$}
\vspace{3mm}

For the limit $p\to 0$, using Eq. (\ref{fouriermetric}), 
we have
\begin{eqnarray}
&&-\frac{1}{2}\Bigl( \Box_{4}-2H^2 \Bigr) h_{\mu\nu}
\nonumber\\
&&\quad=
\frac{1}{2}\int^{\infty}_{-\infty}dp\, Y^{(p,2)}_{\mu\nu} S_{(p,2)}
\frac{\kappa_{5}^2 \ell \sinh(y_0/\ell)H^2(-i p+3/2)
  P^{ip}_{3/2}(z_0)/ 
   P^{ip}_{1/2}(z_0)}
{ 1-\bar{\alpha}
 +\bar{\alpha}(H\ell)^2 \cosh(y_0/\ell)\bigl(-ip+3/2 \bigr)
 P^{ip}_{3/2}(z_0)
 /  P^{ip}_{1/2}(z_0) }
\nonumber\\
&&\quad\simeq 
\frac{3\kappa_{5}^2}{4\ell}
   \frac{(H\ell)P_{3/2}(z_0)/P_{1/2}(z_0)}
{(1-\bar\alpha)
+(3/2)(H\ell)\coth(y_0/\ell)\bar\alpha P_{3/2}(z_0)/P_{1/2}(z_0)}
 \int^{\infty}_{-\infty}dp\,S_{(p,2)} Y^{(p,2)}_{\mu\nu}\,.  
\end{eqnarray}
As for the term involving $\varphi$,
we pull out the part that takes the same form as the above
equation. Using Eq.~(\ref{fourierbendion}), we find
\begin{eqnarray}
&&
2\coth(y_0/\ell) 
\Bigl(D_{\mu}D_{\nu} -\gamma_{\mu\nu}\Box_{4}-3H^2 \gamma_{\mu\nu} 
\Bigr)\varphi
\nonumber\\ 
&&\quad=
\frac{3\kappa_{5}^2}{4\ell}
   \frac{(H\ell)P_{3/2}(z_0)/P_{1/2}(z_0)}
{(1-\bar\alpha)
+(3/2)(H\ell)\coth(y_0/\ell)\bar\alpha P_{3/2}(z_0)/P_{1/2}(z_0)}
 \int^{\infty}_{-\infty}dp\,S_{(p,0)} Y^{(p,0)}_{\mu\nu}
\nonumber
\\
&&\qquad
-\frac{(H\ell)(1-\bar\alpha)P_{-1/2}(z_0)}
{2(1+\beta)P_{1/2}(z_0)
-(H\ell)\coth(y_0/\ell)\bar\alpha P_{-1/2}(z_0)}
\Bigl(D_{\mu}D_{\nu} -\gamma_{\mu\nu}\Box_{4}-3H^2 \gamma_{\mu\nu} 
\Bigr)\varphi\,,
\end{eqnarray}
where we have used the recursion relation,
\begin{eqnarray}
\frac{3}{2}
P_{3/2}(z_0)
=2z_0P_{1/2}(z_0)-\frac{1}{2}P_{-1/2}(z_0)\,.
\end{eqnarray} 
Thus, the effective gravitational equation is expressed as
\begin{eqnarray}
\delta G_{\mu\nu}[\tilde h]
+3H^2\tilde h_{\mu\nu}
&=&\frac{\kappa_{5}^2}{\ell}F_T\,S_{\mu\nu}
-F_S\,
\Bigl(D_{\mu}D_{\nu} -\gamma_{\mu\nu}\Box_{4}-3H^2 \gamma_{\mu\nu} 
\Bigr)\tilde\varphi,
\nonumber
\\
\Bigl(\Box_{4} +4H^2\Bigr) \tilde\varphi
&=&
-\frac{\kappa_5^2}{\ell}\,S\,,
\label{longdis}
\end{eqnarray}
where we have rescaled $\varphi$ to $\tilde\varphi=6(1+\beta)\varphi$,
and $F_T$ and $F_S$ are constants that represent the
tensor and scalar coupling strengths, respectively, given by
\begin{eqnarray}
F_T&=&
   \frac{(H\ell)\Bigl(4\cosh(y_0/\ell)P_{1/2}(z_0)-P_{-1/2}(z_0)\Bigr)}
{2\Bigl(2(1+\beta)P_{1/2}(z_0)-(H\ell)^2\cosh(y_0/\ell)
  \bar\alpha P_{-1/2}(z_0)\Bigr)}\,,
\nonumber
\\
F_S&=&
\frac{(H\ell)(1-\bar\alpha)P_{-1/2}(z_0)}
{6(1+\beta)\Bigl(2(1+\beta)P_{1/2}(z_0)
-(H\ell)^2\cosh(y_0/\ell)\bar\alpha P_{-1/2}(z_0)\Bigr)}\,.
\end{eqnarray}
In the intermediate range of $H\ell$, i.e., when $H\ell=O(1)$, then 
$F_T$ and $F_S$ are comparable and
we obtain a Brans-Dicke type theory given by Eq. (\ref{fluctuation})
with the identifications,
\begin{eqnarray}
&&\frac{8\pi G_{4}}{ \Phi_{0}}
=\frac{\kappa_5^2}{\ell}F_T \,,
\quad
\frac{\delta\Phi}{\Phi_{0}}=-F_{S} \tilde{\varphi}\,,
\quad
\Lambda_{4}=3H^2\,,
\\ \nonumber
&&\omega=\frac{F_T-3F_S}{2 F_S}
 =\frac{6(1+\beta)\cosh(y_0/\ell)P_{1/2}(z_0)
        -3\Bigl(1+(H\ell)^2\bar{\alpha}\Bigr)P_{-1/2}(z_0)}
       {(1-\bar{\alpha})P_{-1/2}(z_0)}.
\end{eqnarray} 
A potential problem in this
case is that the tachyonic mass of the scalar field seems to make
the system unstable. However, as discussed in Sec.~III~D,
the tachyonic pole is not excited by the matter source.
Further, as discussed in Sec.~III~E, the source-free tachyonic
mode do not cause an instability either.

For $H\ell\ll1$,
$\omega \gg 1$ and
the scalar field decouples to yield
\begin{eqnarray}
&& \delta G_{\mu\nu}[\tilde h]
+3H^2\tilde h_{\mu\nu}
=\frac{\kappa_{5}^2}{\ell}
  \frac{\coth(y_0/\ell)}{ 1+\beta}S_{\mu\nu}.
\end{eqnarray}
Thus we obtain the Einstein gravity with
\begin{eqnarray}
 8\pi G_{4}=\frac{\kappa_{5}^2}{\ell}
        \frac{\coth(y_0/\ell)}{1+\beta}\,.
\end{eqnarray}
In the limit $H\ell\to0$,
\begin{eqnarray}
 8\pi G_{4}\simeq 
   \frac{\kappa_{5}^2}{\ell}\frac{1}{1+\bar{\alpha}}.
\end{eqnarray}
This is the result for the Minkowski brane.

In the limit $H\ell\gg1$, $\omega \gg 1$ and we recover the
4-dimensional Einstein gravity on the brane with
\begin{eqnarray}
8\pi G_4=\frac{\kappa_5^2}{2(H\ell)\bar\alpha\ell}\,.
\end{eqnarray}
Note that this is just a special case of the high energy brane
case discussed in subsection~A.

Thus we conclude that
despite the presence of the tachyonic mass, the system is
stable and well-behaved for all ranges of $H\ell$.

\section{Summary and Discussion}

We have investigated the linear perturbation of a de Sitter brane
in an Ant-de Sitter bulk in the 5-dimensional Einstein Gauss-Bonnet
(EGB) theory. We have derived the effective theory on the brane
which is described by a set of integro-differential equations. 

To understand the nature of this theory in more details, we have
investigated the behavior of the theory in various limiting cases.
In contrast to the case of a braneworld in
the 5-dimensional Einstein gravity,
in which both the short distance and high energy brane limits 
exhibit 5-dimensional behavior, we have found that the gravity
on the brane is effectively 4-dimensional for all the limiting cases.

For a high energy brane, i.e., in the limit $H\ell\gg1$,
the Einstein gravity is recovered, provided that
the length scale of fluctuations is of order $H^{-1}$.
It is found that the low energy corrections are suppressed by
the factor $O\bigl((H\ell)^{-2}\bigr)$.

In the short distance limit $r\ll \min\{\ell,H^{-1}\}$,
the scalar field that describes brane bending becomes dynamical,
and we obtain the Brans-Dicke gravity. This is consistent
with the case of the Minkowski brane. A slight complication
is that this brane-bending scalar
field is tachyonic, with mass-squared $-4H^2$. Therefore,
if it becomes dynamical, one would naively expect the theory
to become unstable. However, since the energy scale
of fluctuations are much larger than $H$, the fluctuations
actually do not see this tachyonic mass, hence there is
no instability.

In the large distance limit $r\gg \max\{\ell,H^{-1}\}$, the Einstein gravity
is obtained in both limits $H\ell\ll1$ and $H\ell\gg1$,
while a Brans-Dicke type theory is obtained for $H\ell=O(1)$.
Although the scalar field of this Brans-Dicke gravity is tachyonic with
mass-squared given by $-4H^2$,
we have shown that this mode is not excited by the matter
source, hence does not lead to an instability of
the system.

In the limit $H\ell\to0$, the previous results for
the Minkowski brane have been recovered, that is,
the Brans-Dicke gravity at short distances and
the Einstein gravity at large distances.

In all the cases, the effective 4-dimensional gravitational
constant depends non-trivially on the values of $H\ell$ and
$\bar\alpha$, where $\bar\alpha$ is the non-dimensional
coupling constant for the Gauss-Bonnet term. This indicates
the time variation of the gravitational constant in the course
of the cosmological evolution of a brane in the EGB theory.
It will be interesting to investigate
in more details the cosmological implications of the 
braneworld in the EGB theory.

\section*{Acknowledgments}

We are grateful to S.~Kanno and J.~Soda 
for invaluable discussions. Discussions with them,
especially at the early stage of this work, gave us
important physical insights into the behavior of
the EGB braneworld.
This work was supported in part by Monbu-Kagakusho
 Grant-in-Aid for Scientific Research (S) No.\ 14102004.

\appendix

\section{The results for the Minkowski brane}

Here, we summarize the results for the Minkowski brane~\cite{Davis:2004xa}.

\subsection{Effective equations on the brane}

In the RS gauge, the perturbed metric in the bulk is
written
\begin{eqnarray}
ds^2= dy^2+b^2(y) \bigl(\eta_{\mu\nu}+h_{\mu\nu}\bigr)dx^{\mu}dx^{\nu},
\quad b(y)=e^{-|y|/\ell},
\end{eqnarray}
where $\eta_{\mu\nu}$ is the Minkowski metric.
The brane locates at $y=0$ in the background.  
The background part of the Einstein Gauss-Bonnet equation~(\ref{fullEGB})
gives the relation of the AdS radius to the bulk cosmological constant,
Eq.~(\ref{AdS}).
The perturbative part of Eq.~(\ref{fullEGB}) gives 
\begin{eqnarray}
\Bigl(1-\bar{\alpha}\Bigr)
\Bigl(\partial_{y}^2-4\frac{1}{\ell}\partial_{y}+e^{2y/\ell}\Box_4
\Bigr)h_{\mu\nu} =0\,.
\end{eqnarray}
Again, we consider the case $\bar{\alpha}\neq 1$.
The location of the brane is perturbed to be at
$y=-\ell\varphi$. 
Induced metric on the brane is given by
\begin{eqnarray}
ds^2 \Big|_{(4)}= \Bigl(\eta_{\mu\nu}+\tilde h_{\mu\nu} \Bigr)
                 dx^{\mu}dx^{\nu},\quad
\tilde h_{\mu\nu}=h_{\mu\nu}-2\varphi \eta_{\mu\nu}\,.
\end{eqnarray}

The solution for $h_{\mu\nu}$ on the brane which satisfies
the junction condition is given by
\begin{eqnarray}
 h_{\mu\nu}\Big|_{y=0}
  =-\frac{\kappa_{5}^2}{\ell}\int\frac{d^4p}{(2\pi)^4}e^{ip\cdot x}
  \frac{\ell^2 H^{(1)}_2(q\ell)}{(1-\bar\alpha)q\ell H^{(1)}_1(q\ell)
  +\bar\alpha q^2\ell^2H^{(1)}_2(q\ell)}\left[
  S_{\mu\nu}(p)-\frac{1}{3}
  \left(\eta_{\mu\nu}-\frac{p_\mu p_\nu}{p^2}\right)
   S(p)\right],
\label{fire}
\end{eqnarray}
where $H^{(1)}_{\nu}$ is the Hankel function of the first kind
 and $q^2=-p^2$. The equation that determines the brane
bending is 
\begin{eqnarray}
\Box_4\,\varphi=-\frac{\kappa_{5}^2}{6\ell}\frac{1}{1+\bar\alpha}S.
\label{BB}
\end{eqnarray}

The perturbed 4-dimensional Einstein tensor is expressed as
\begin{eqnarray}
\delta G_{\mu\nu}[\tilde h]=-\frac{1}{2}\Box_4\, h_{\mu\nu}
 +2\left(\partial_\mu\partial_\nu
  -\eta_{\mu\nu}\Box_4\,\right)\varphi.
\label{einstein}
\end{eqnarray}
Inserting Eq.~(\ref{fire}) into Eq.~(\ref{einstein}), we 
obtain the effective equation on the brane, which reads
\begin{eqnarray}
\delta G_{\mu\nu}[\tilde h]&=&
 \frac{\kappa_{5}^2}{2\bar\alpha\ell}
\int\frac{d^4p}{(2\pi)^4}e^{ip\cdot x} 
 \frac{\bar\alpha q^2\ell^2 H^{(1)}_2(q\ell)}
 {(1-\bar\alpha)q\ell H^{(1)}_1(q\ell)
 +\bar\alpha q^2\ell^2H^{(1)}_2(q\ell)}
 \left[
 S_{\mu\nu}(p) -\frac{1}{3}
 \left(\eta_{\mu\nu}-\frac{p_\mu p_\nu}{p^2}\right)
 S(p)\,\right]\nonumber\\
&&
+2\left(\partial_\mu\partial_\nu
-\eta_{\mu\nu}\Box_4\,\right)\varphi.
\label{effG}
\end{eqnarray}

\subsection{Short distance limit}

In the short distance limit $q\ell\gg1$,
 Eq.~(\ref{effG}) becomes
\begin{eqnarray}
\delta G_{\mu\nu}[\tilde h]&=&
\frac{\kappa_{5}^2}{2\bar\alpha\ell}S_{\mu\nu}
 -\left(\frac{1-\bar\alpha}{\bar\alpha}\right)
 \left(\partial_\mu\partial_\nu
 -\eta_{\mu\nu}\Box_4\,\right)\varphi.
\label{Minshort}
\end{eqnarray}
Comparing Eqs.~(\ref{BB}) and (\ref{Minshort})
 with the linearized Brans-Dicke gravity,
\begin{eqnarray}
\delta G_{\mu\nu}[\tilde h]=\frac{8\pi G_4}{\Phi_0}
S_{\mu\nu}
+\frac{1}{\Phi_0}
\left(\partial_\mu\partial_\nu
-\eta_{\mu\nu}\Box_4\,\right)\delta\Phi\,,\quad
\Box_4\,\delta \Phi=\frac{8\pi G_4}{3+2\omega}S,
\label{Brans-Dicke}
\end{eqnarray}
we find the correspondences,
\begin{eqnarray}
\frac{8\pi G_4}{\Phi_0}=\frac{\kappa_{5}^2}{2\bar{\alpha}\ell},
\quad
\frac{\delta \Phi}{\Phi_0}= -\frac{1-\bar{\alpha}}{\bar{\alpha}}\varphi,
\quad  
\omega=\frac{3\bar\alpha}{1-\bar\alpha}.
\end{eqnarray}

The corrections are rewritten as
\begin{eqnarray}
\Bigl( \delta G_{\mu\nu}[\tilde h]\Bigr)_{\mathrm{corr}}
    =-\frac{\kappa_{5}^2}{2\bar\alpha\ell}
\int\frac{d^4p}{(2\pi)^4}e^{ip\cdot x} 
\frac{(1-\bar\alpha)q\ell H^{(1)}_1(q\ell)}
{(1-\bar\alpha)q\ell H^{(1)}_1(q\ell)
+\bar\alpha q^2\ell^2H^{(1)}_2(q\ell)}
\left[
S_{\mu\nu} -\frac{1}{3}
\left(\eta_{\mu\nu}-\frac{p_\mu p_\nu}{p^2}\right)
 S\,\right].
\end{eqnarray}

\subsection{Large distance limit}

In the large distance limit $q\ell\ll 1$, 
Eq.~(\ref{effG}) becomes
\begin{eqnarray}
\delta G_{\mu\nu}[\tilde h]=\frac{\kappa_{5}^2}{\ell}
\frac{1}{1+\bar\alpha}S_{\mu\nu}.
\end{eqnarray}
Thus we obtain the Einstein gravity with
\begin{eqnarray}
8\pi G_{4}=\frac{\kappa_{5}^2}{\ell}\frac{1}{1+\bar\alpha}.
\end{eqnarray}

\section{Harmonic Functions on de Sitter geometry}

In this Appendix, we consider the harmonics on the de Sitter
 spacetime with curvature radius $H^{-1}$.
They are obtained by the Lorentzian generalization
of the tensor harmonics on an $n$-dimensional constant curvature 
Riemannian space~\cite{Kodama:bj}.
We focus on the tensor-type and scalar-type harmonics.

\subsection{Tensor-type harmonics}

The tensor-type tensor harmonics satisfy
\begin{eqnarray}
\Bigl(\Box_{4}-\bigl(p^2+17/4\bigr)
 H^2 \Bigr)Y^{(p,2)}_{\mu\nu}(x^{\mu})=0,
\end{eqnarray}
which corresponds to the 4-dimensional massive gravitons with
mass-squared $m^2H^2=(p^2+9/4)H^2$.
They satisfy the transverse-traceless condition,
\begin{eqnarray}
Y^{(p,2)\mu}{}_{\mu}=Y^{(p,2)}_{\quad\mu}{}^{\nu}{}_{|\nu}=0.
\end{eqnarray}

In reality, the tensor harmonics have 3 more
indices for the spatial eigenvalues.
If we adopt the flat slicing,
\begin{eqnarray}
ds^2=-dt^2+H^{-2}e^{2Ht}\delta_{ij}dx^i dx^j\,,
\end{eqnarray}
we can use the standard Fourier modes $e^{i\bm{k}\cdot\bm{x}}$,
and the spatial indices will be continuous.
In addition, we also have discrete indices $\sigma$
that describe the polarization degrees of freedom
(5 in 4-dimensions).
However, for notational simplicity, we omit these indices.

We ortho-normalize the tensor harmonics as
\begin{eqnarray}
\int d^4 x \sqrt{-\gamma}\, 
Y^{(p,2)}_{\mu\nu}\,Y^{*(p',2)}{}^{\mu\nu} 
     = \delta(p-p')\delta^3( \bm{k}-\bm{k}')\delta_{\sigma,\sigma'}\,.
\end{eqnarray} 
Although we have no explicit proof for the completeness, due to
our poor knowledge, we assume that $Y^{(p,2)}_{\mu\nu}$ 
for $-\infty<p<\infty$ constitute a complete set
for the space of transverse-traceless tensors.

\subsection{Scalar-type harmonics}

The scalar-type harmonics $Y^{(p,0)}(x^{\mu})$
satisfy the equation for a scalar field with mass-squared 
$m^2H^2=(p^2+9/4)H^2$,
\begin{eqnarray}
\Bigl(\Box_{4}-\bigl(p^2+\frac{9}{4} \bigr)
 H^2 \Bigr)Y^{(p,0)}(x^{\mu})=0.
\end{eqnarray} 
We assume they satisfy the ortho-normality condition,
\begin{eqnarray}
\int d^4 x \sqrt{-\gamma}\, Y^{(p,0)}Y^{*(p',0)} 
= \delta(p-p')\delta^3( \bm{k}-\bm{k'}).
\end{eqnarray}

{}From $Y^{(p,0)}$, the ortho-normalized
scalar-type vector harmonics are constructed as
\begin{eqnarray}
Y^{(p,0)}_{\mu}= \frac{i}{H\sqrt{p^2+9/4} }D_{\mu} Y^{(p,0)},
\end{eqnarray}
which satisfy
\begin{eqnarray}
\int d^4 x \sqrt{-\gamma}\, Y^{(p,0)}_{\mu}Y^{*(p',0)\mu} 
= \delta(p-p') \delta^3( \bm{k}-\bm{k'}) .
\end{eqnarray}

The trace-free and divergence-free scalar-type tensor harmonics
are constructed, respectively, as
\begin{eqnarray}
&&\bar{Y}^{(p,0)}_{\mu\nu}
=N_{p}\Bigl[D_{\mu}D_{\nu} Y^{(p,0)}
 -\frac{1}{4}\Bigl(p^2+\frac{9}{4}\Bigr)
\gamma_{\mu\nu}H^2 Y^{(p,0)}
 \Bigr] 
\nonumber\\
&&Y^{(p,0)}_{\mu\nu}
=N_{p}\Bigl[ D_{\mu}D_{\nu} Y^{(p,0)}
 -\Bigl( p^2+\frac{21}{ 4}\Bigr) \gamma_{\mu\nu}H^2 Y^{(p,0)}
 \Bigr] 
\nonumber\\
&& \hspace{1cm}
=\bar{Y}^{(p,0)}_{\mu\nu}
-\frac{3}{4}N_{p} \Bigl(p^2+\frac{25}{ 4}\Bigr)H^2\gamma_{\mu\nu}
Y^{(p,0)}, 
\end{eqnarray}
where
\begin{eqnarray}
|N_{p}|^2= \frac{1}{ 3\bigl(p^2+21/4\bigr)\bigl(p^2+25/4\bigr)H^4}, 
\end{eqnarray}
Without loss of generality, we assume that 
$N_p$ is real and positive. 
The scalar-type divergence-free tensor harmonics
$Y^{(p,0)}_{\mu\nu}$ satisfy the ortho-normality condition,
\begin{eqnarray}
\int d^4 x \sqrt{-\gamma}\, Y^{(p,0)}_{\mu\nu}
                            Y^{*(p',0)}{}^{\mu\nu} 
=\delta(p-p')\delta^3 ( \bm{k}-\bm{k}') .
\end{eqnarray}

\section{Tachyonic bound state in de Sitter two-brane system}

 In \cite{Charmousis:2003sq}, Charmousis and Dufaux showed that
for the Minkowski two-brane system there exists
 a tachyonic bound state on the negative tension brane.
This fact implies that the Minkowski two-brane system  
is unstable under the linear perturbation.
Following~\cite{Charmousis:2003sq}, we show that there exits a 
tachyonic bound state
also for the de Sitter two-brane system.

\subsection{Possibility of a negative norm state}

We consider a de Sitter two-brane system. One of
the branes located at a smaller radius of the AdS space
has a negative tension.
 We discuss only the bulk gravitational perturbations.
The matter perturbations on each brane are not taken into account.

The bulk component of the perturbed Einstein Gauss-Bonnet equation
including the boundary branes are written in the Sturm-Liouville form as
\begin{eqnarray}
\Bigl\{ \Bigl( b^4-\bar\alpha\,\ell^2
 \bigl(b^2b'^2-b^2 H^2 \bigr)\Bigr)\psi_{p,y}\Bigr\}_{,y}
 =-b^2\Bigl(1-\bar\alpha\,\ell^2 \frac{b''}{ b} \Bigr)
\Bigl(p^2+\frac{9}{4}\Bigr) H^2\psi_{p}. \label{EGBapp}
\end{eqnarray} 
Using Eq. (\ref{EGBapp}), the boundary condition on each brane is derived.
For $H=0$ and $b(y)=e^{-|y|/\ell}$, Eq.~(\ref{EGBapp}) naturally
reduces to the Minkowski version, Eq. (8) in \cite{Charmousis:2003sq}.

\begin{enumerate}

\item{{\bf On positive tension brane}}

Imposing the $Z_{2}$ symmetry, the warp factor around the
positive tension brane is expressed as
\begin{eqnarray}
b(y)=H\ell\, \sinh\Bigl(\frac{y_{+}-|y-y_{+}|}{\ell} \Bigr).
\end{eqnarray}
Integrating Eq.~(\ref{EGBapp}) around $y=y_+$ and using the
$Z_2$-symmetry,
\begin{eqnarray}
\partial_y\psi_{p}(y_{+}-0)
= \frac{\zeta}{\ell} 
\frac{(p^2+9/4)\cosh(y_{+}/\ell)}{\sinh^3(y_{+}/\ell)}
             \,\psi_{p}(y_+),  
\label{pos}
\end{eqnarray}
where
\begin{eqnarray}
\zeta:= \frac{\bar\alpha}{1-\bar\alpha}\,.
\end{eqnarray}

\item{{\bf On negative tension brane}}

Similarly, the $Z_{2}$ symmetry gives the warp factor
around the negative tension brane as
\begin{eqnarray}
b(y)=H\ell\, \sinh\Bigl(\frac{|y-y_{-}|+y_{-}}{\ell} \Bigr).
\end{eqnarray}
Integrating Eq.~(\ref{EGBapp}) around $y=y_-$ and using the $Z_2$-symmetry,
\begin{eqnarray}
\partial_y\psi_{p}(y_{-}+0)
= \frac{\zeta }{\ell} 
  \frac{(p^2+9/4)\cosh({y_{-}}/\ell ) }
 {\sinh^3(y_{-}/\ell)}\,\psi_{p}(y_-).
\label{neg} 
\end{eqnarray}

\end{enumerate}

For both branes, the boundary conditions are of a mixed (Robin) type.
This renders us impossible to prove the positivity of the norm.
Namely, we have 
\begin{eqnarray}
&&
\int_{y_-}^{y_+}dy\,
 \Bigl( b^4-\bar\alpha\,\ell^2
\bigl(b^2b'^2-b^2 H^2\bigr) \Bigr)(\partial_y\psi_{p})^2   
\nonumber\\
&&=\bigl(H\ell\bigr)^4\bigl(p^2+\frac{9}{4} \bigr)
\Biggl[\frac{\bar\alpha}{2 \ell}
\Bigl(\sinh(2{y_+}/{\ell})\psi_{p}^2(y_{+})
    -\sinh(2{y_-}/{\ell})\psi_{p}^2(y_{-})\Bigr)
+\frac{(1-\bar\alpha)}{ \ell^2}
\int_{y_-}^{y_+}dy\,
 \sinh^2\bigl(y/\ell\bigr) \psi_{p}^2(y)\Biggr].  
\end{eqnarray}
Thus the norm is no longer positive definite for $p^2+9/4>0$.

\subsection{Condition for the existence of tachyonic bound state}

In order to determine whether a tachyonic bound state exists, 
we need to analyze the mass spectrum.
The tachyonic eigenmode, if it exists, is written by
\begin{eqnarray}
\psi_{q}(y)
=\frac{1}{\sinh^{3/2}\bigl(y/\ell\bigl)}
\Bigl[A_q P^{-q}_{3/2}\bigl(\cosh\bigl(y/\ell\bigr) \bigr)
+B_q P^{q}_{3/2}\bigl(\cosh\bigl(y/\ell\bigr) \bigr)
   \Bigr],
\end{eqnarray}
where $m^2=-\mu^2$, $q:= \sqrt{\mu^2+9/4}$, and $q^2=-p^2$.
The $y$-derivative of it is
\begin{eqnarray}
\partial_y\psi_{q}
=-\frac{1}{\ell\sinh^{5/2}\bigl({y}/{\ell}\bigl)}
\Bigl[
\bigl(\frac{3}{2}-q\bigr)
A_q P^{-q}_{1/2}\bigl(\cosh\bigl(y/\ell\bigr) \bigr)
+\bigl(\frac{3}{2}+q\bigr)
B_q P^{q}_{1/2}\bigl(\cosh\bigl(y/\ell\bigr) \bigr)
   \Bigr].
\end{eqnarray}

Using the boundary condition on each brane,
Eqs.~(\ref{pos}) and (\ref{neg}), we obtain
\begin{eqnarray}
&&A_q \bigl(\frac{3}{2}-q \bigr)
 \Bigl((z_+^2-1) P^{-q}_{1/2}(z_+)
+\zeta \bigl(\frac{3}{2}+q \bigr)z_+ 
P^{-q}_{{3}/{2}}(z_+)
\Bigr) 
 \nonumber\\
&&\quad
+B_q \bigl(\frac{3}{2}+q \bigr)
\Bigl((z_+^2-1) P^{q}_{{1}/{2}}(z_+)
+\zeta  \bigl(\frac{3}{2}-q \bigr)z_+
 P^{q}_{{3}/{2}}(z_+)
\Bigr)=0, 
 \nonumber\\
&&
A_q \bigl(\frac{3}{2}-q \bigr)
 \Bigl((z_-^2-1) P^{-q}_{1/2}(z_-)
+\zeta \bigl(\frac{3}{2}+q \bigr)z_-
P^{-q}_{3/2}(z_-)
\Bigr) 
 \nonumber\\
&&\quad
+B_q \bigl(\frac{3}{2}+q \bigr)
\Bigl((z_-^2-1) P^{q}_{{1}/{2}}(z_-)
+\zeta  \bigl(\frac{3}{2}-q \bigr)z_-
 P^{q}_{3/2}(z_-)
\Bigr)=0, 
\end{eqnarray}
where $z_{\pm}=\cosh(y_{\pm}/\ell)$. 

For a non-trivial solution for $A_q$ and $B_q$ to exist,
the determinant must vanish. Thus 
\begin{eqnarray}
&&
\Bigl( (z_+^2-1)P^{-q}_{1/2}(z_+)
+\zeta \bigl(\frac{3}{2}+q \bigr)z_+ 
P^{-q}_{{3}/{2}}(z_+)
\Bigr) 
\Bigl((z_-^2-1)  P^{q}_{1/2}(z_-)
+\zeta  \bigl(\frac{3}{2}-q \bigr)z_- 
 P^{q}_{{3}/{2}}(z_-)
\Bigr)
\nonumber \\
&&
-\Bigl((z_-^2-1) P^{-q}_{1/2}(z_-)
+\zeta \bigl(\frac{3}{2}+q \bigr)z_- 
P^{-q}_{{3}/{2}}(z_-)
\Bigr)
 \Bigl( ( z_+^2-1) P^{q}_{1/2}(z_+)
+\zeta \bigl(\frac{3}{2}-q \bigr)z_+ 
P^{q}_{{3}/{2}}(z_+)
\Bigr)
=0 .
\label{determinant}
\end{eqnarray}
The pole at $q=3/2$, which corresponds to the zero mode, is
divided out in deriving Eq.~(\ref{determinant}).  
If there exists a solution of Eq.~(\ref{determinant}) at $q>3/2$,
it implies the existence of a tachyonic bound state.

\subsection{Existence of a tachyonic bound state}

{}From Eq.~(\ref{determinant}),
\begin{eqnarray}
\frac{ (z_-^2-1)P^{q}_{1/2}(z_-)
+\zeta \bigl(\frac{3}{2}-q \bigr)z_- 
P^{q}_{{3}/{2}}(z_-)} 
{(z_-^2-1)  P^{-q}_{1/2}(z_-)
+\zeta  \bigl(\frac{3}{2}+q \bigr)z_- 
 P^{-q}_{{3}/{2}}(z_-)}
=
\frac{(z_+^2-1) P^{q}_{1/2}(z_+)
+\zeta \bigl(\frac{3}{2}-q \bigr)z_- 
P^{q}_{{3}/{2}}(z_-)}
 { ( z_+^2-1) P^{-q}_{1/2}(z_+)
+\zeta \bigl(\frac{3}{2}+q \bigr)z_+ 
P^{-q}_{{3}/{2}}(z_+)}\,. \label{determinant2}
\end{eqnarray}
Using the definition of the Legendre functions~\cite{Bateman:1969ku},
\begin{eqnarray}
P_{\nu}^{\mu}(z)
= \frac{1}{\Gamma(1-\mu)} \Bigl(\frac{z+1}{z-1}\Bigr)^{\mu/2}
{}_2F_1\Bigl[-\nu,\,\nu+1;\,1-\mu;\,\frac{1-z}{2}\Bigr],
\end{eqnarray}
we see that the left-hand-side of Eq.~(\ref{determinant2}) is generally
much larger than the right-hand-side for $q\gg 1$ for fixed $z_{+}$
and $z_{-}$. Therefore, in order for this equation to be satisfied, 
we must have
\begin{eqnarray}
q-\frac{3}{2}\simeq
\frac{(z_-^2-1)P_{1/2}^{q}(z_-) }{\zeta z_- P_{3/2}^{q}(z_-)}
\to
\frac{z_-^2-1 }{\zeta\, z_-}
\quad\mbox{for}~ q\to\infty\,.
\end{eqnarray}
This is a consistent solution for $\zeta\ll1$. Thus a tachyonic bound
state exists in the de Sitter brane case as well.

The tachyon mass is given by
\begin{eqnarray}
\mu H=\sqrt{q^2-9/4}\,H \simeq  \frac{(z_-^2-1)H\ell}{\zeta z_{-}\,\ell}\,.
\end{eqnarray}
In the low energy limit, we have $z_{+}>z_{-}\gg1$ and
$H\ell\simeq1/z_{+}\ll1$. Hence, the above reduces to
\begin{eqnarray}
\mu H \simeq \frac{\Omega}{\zeta\, \ell}\,,
\end{eqnarray}
where 
\begin{eqnarray}
\Omega :=\frac{b(z_-)}{b(z_+)}\simeq
\frac{z_{-}}{z_{+}}\sim e^{-(y_+ -y_-)/\ell}\,.
\end{eqnarray}
This agrees with the result for the Minkowski brane~\cite{Charmousis:2003sq}.

On the other hand, in the high energy limit, $H\ell \gg 1$, we have
\begin{eqnarray}
\mu H\simeq \frac{\Omega^2 H}{\zeta (H\ell)^2}\ll \frac{H}{\zeta}\,. 
\end{eqnarray}
Thus the high background expansion rate of the brane suppresses 
the tachyonic mass, giving a tendency to stabilize the two-brane
system.

\end{document}